\documentclass[twocolumn,showpacs]{revtex4}
\usepackage{graphicx}
\usepackage{dcolumn}

\begin{document}


\title{Quenching of pairing gap at finite temperature in $^{184}$W}

\author{K. Kaneko$^{1}$ and M. Hasegawa$^{2}$}
\affiliation{
$^{1}$Department of Physics, Kyushu Sangyo University, Fukuoka 813-8503, Japan \\
$^{2}$Laboratory of Physics, Fukuoka Dental College, Fukuoka 814-0193, Japan 
}

\date{\today}

\begin{abstract}

We extract pairing gap in $^{184}$W at finite temperature for the first time 
from the experimental level densities of $^{183}$W, $^{184}$W, and $^{185}$W using ``thermal" 
odd-even mass difference. We found the quenching of pairing gap near the critical temperature 
$T_c = 0.47$ MeV in the BCS calculations. 
It is shown that the monopole pairing model with a deformed Woods-Saxon potential explains 
the reduction of the pairing correlation using the partition function with the number 
parity projection in the static path approximation plus random-phase approximation. 

\end{abstract}

\pacs{21.60.Jz, 21.10.Ma, 05.30.-d}

\maketitle

Pairing correlations are essential for many-fermion systems such as electrons in the 
superconducting metal, nucleons in the nucleus, and quarks in the color superconductivity. 
The Bardeen-Cooper-Schriffer (BCS) theory \cite{BCS} of superconductivity has succesfully 
described the pairing correlations. This theory was applied to nuclear problems at zero 
temperature \cite{Bohr,Belyaev} for stable nuclei. The thermodynamical properties of nuclear 
pairing were investigated by using the BCS theory in the study of hot nuclei \cite{Sano,Goodman}. 
Breaking of the Cooper pairs is expected to occur at a certain critical temperature in the 
BCS theory. 

It has recently been reported \cite{Schiller,Melby} that the canonical heat capacities 
extracted from the observed level densities in $^{162}$Dy, $^{166}$Er and $^{172}$Yb form the S 
shape with a peak around the temperature $T\approx 0.5$ MeV. 
These S-shaped heat capacities were interpreted as the 
breaking of nucleon Cooper pairs and the pairing transition because this temperature is close 
to the critical temperature $T_{c} = 0.57\Delta \approx 0.5$ MeV in the BCS theory. 
For the finite Fermi system like a nucleus, however, nucleon number fluctuation 
and statistical fluctuations 
beyond the mean field become large. The fluctuations wash out the sharp phase transition, and 
then the pairing gap $\Delta$ does not become quickly zero at the BCS critical temperature. 
Several models have taken into account the fluctuations beyond the mean field. 
The quenching of pairing correlations have been 
obtained in recent theoretical approaches: 
the static path approximation (SPA) plus random-phase approximation 
(RPA) \cite{Rossignoli}, 
the shell model Monte Calro (SMMC) calculations \cite{Rombouts,Liu}, 
the finite-temperature Hartree-Fock-Bogoliubov (HFB) theory \cite{Egido}, 
and the relativistic mean field theory \cite{Agrawal}. 

The odd-even mass difference observed in nuclear masses is well known as one of signatures 
of pairing correlations. 
In solid state physics, difference between the free energies with odd and even 
numbers of electrons in ultrasmall superconducting grains is found and known as even-odd effect 
\cite{Tuominen}. 
In our previous paper \cite{Kaneko}, we suggested that the suppression of the pairing 
correlations due to finite temperature appears in ``thermal" odd-even mass difference 
rather than the S shape of the heat capacity. 

In this Rapid Communication, using the thermal odd-even mass difference we extract the 
pairing gap of $^{184}$W from the experimental level densities of $^{183}$W, $^{184}$W, 
and $^{185}$W recently observed \cite{Bondarenko,Sukhovoj}. 
It is shown that the reduction of the thermal odd-even mass difference is interpreted 
as a signature of pairing transition, but not the S shape of the heat capacity. 

\begin{figure}[t]
\includegraphics[width=8cm,height=10cm]{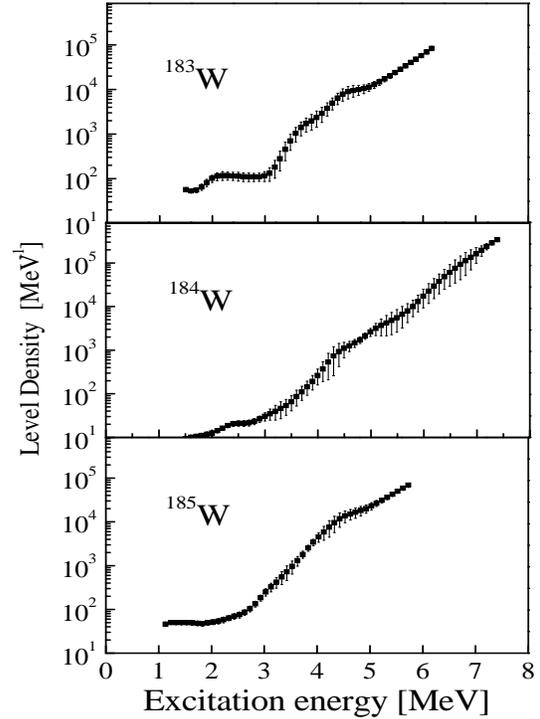}
  \caption{Experimental level densities for $^{183}$W, $^{184}$W, and $^{185}$W. 
  The error bars show the statistical uncertainties. }
  \label{fig1}
\end{figure}

Figure \ref{fig1} shows the experimental level densities of $^{183}$W, $^{184}$W, and $^{185}$W 
extracted from the two-step $\gamma$-cascade intensities. 
To study the thermal properties from the measured level densities, 
let us start from the partition function in the canonical ensemble with the Laplace 
transform of the level density $\rho(E_{i})$ 
\begin{eqnarray}
Z(T) = \sum_{i=0}^{\infty}\delta E_{i}\rho(E_{i}){\rm e}^{-E_{i}/T}, 
\label{eq:1}
\end{eqnarray}
where $E_{i}$ are the excitation energies and $\delta E_{i}$ are the energy bins. 
Then any thermodynamical quantities $O(Z,N,T)$ can be evaluated by 
\begin{eqnarray}
O(Z,N,T) = \sum_{i=0}^{\infty}\delta E_{i}\rho(E_{i})O_{i}{\rm e}^{-E_{i}/T}/Z(T). 
\label{eq:2}
\end{eqnarray}
For instance, the thermal energy is expressed as
\begin{eqnarray}
E(Z,N,T) = \sum_{i=0}^{\infty}\delta E_{i}\rho(E_{i})E_{i}{\rm e}^{-E_{i}/T}/Z(T). 
\label{eq:3}
\end{eqnarray}
The heat capacity are then given by 
\begin{eqnarray}
C(Z,N,T) & = & \frac{\partial E(Z,N,T)}{\partial T}. 
\label{eq:4}
\end{eqnarray}

We now introduce the thermal odd-even mass difference for neutrons defined by 
the following three-point indicator: 
\begin{eqnarray}
\Delta_{n}^{(3)}(Z,N,T) & = & \frac{(-1)^{N}}{2}[ E_{s}(Z,N+1,T) \nonumber \\
   & {} & -2E_{s}(Z,N,T)+E_{s}(Z,N-1,T)], \nonumber \\
\label{eq:5}
\end{eqnarray}
where $E_{s}$ is a shifted thermal energy which is defined 
by substracting the the Coulomb energy from the binding energy at zero temperature. 
The odd-even mass difference at zero temperature is known theoretically and experimentally 
as an important quantity to evaluate the pairing correlations in a nucleus. The thermal 
odd-even mass difference would be also an indicator of the pairing correlations at 
finite temperature, and is obtained from the experimental energies and the level 
density as well as the heat capacity. 

\begin{figure}[b]
\includegraphics[width=8cm,height=10cm]{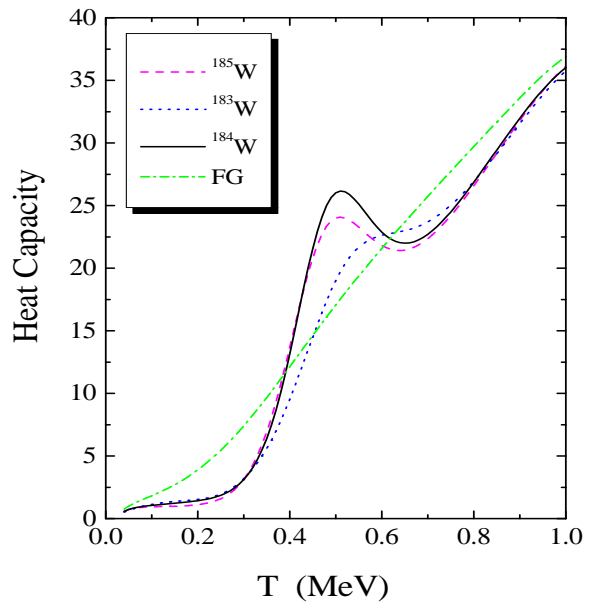}
  \caption{Heat capacities as a function of temperature $T$. 
  The broken, dotted, and solid lines denote, respectively, those of $^{183}$W, 
  $^{184}$W, and $^{185}$W. The broken dotted line indicates the heat capacity 
  of the back-shifted level density formula.}
  \label{fig2}
\end{figure}

We can calculate the canonical partition function $Z(T)$ and the thermodynamical 
quantities from the measured level densities. 
Formally, the calculations using Eqs. (\ref{eq:1})-(\ref{eq:3}) require infinite summation. 
However, the experimental level densities of Fig. \ref{fig1} only cover the excitation energy 
up to $6-8$ MeV. In the evaluation of Eqs. (\ref{eq:1})-(\ref{eq:5}), therefore, 
we extrapolate 
the plots of the experimental density to $\sim$ 40 MeV. Here, we use the level density 
formula of the back-shifted Fermi gas model in Refs. \cite{Gilbert} 
\begin{eqnarray}
\rho(U) & = & f\frac{{\rm exp}[2\sqrt{aU}]}{12\sqrt{2}a^{1/4}U^{5/4}\sigma}, 
\label{eq:6}
\end{eqnarray}
where the back-shifted energy is $U=E-E_{1}$ and the spin cutoff parameter $\sigma$ 
is defined through $\sigma^{2}=0.0888A^{2/3}\sqrt{aU}$. The level density parameter 
$a$ and the back-shifted parameter $E_{1}$ are defined by $a=0.21A^{0.87}$ MeV$^{-1}$ 
and $E_{1}=C_{1}+\Delta$, respectively, where the correction factor is given by 
$C_{1}=-6.6A^{-0.32}$. The factor $f$ is determined 
so as to adjust the back-shifted level density to experimental one. 
The factors are, respectively, 0.3, 0.4, and 0.7 for $^{183}$W, $^{184}$W, and $^{185}$W. 
The parameters $\Delta$ for $^{183}$W and $^{185}$W are taken as the neutron 
pairing energies 0.61 and 0.72 MeV deduced from mass differences, respectively, 
and for $^{184}$W it is fixed at 1.60 MeV. 

\begin{figure}[t]
\includegraphics[width=8cm,height=10cm]{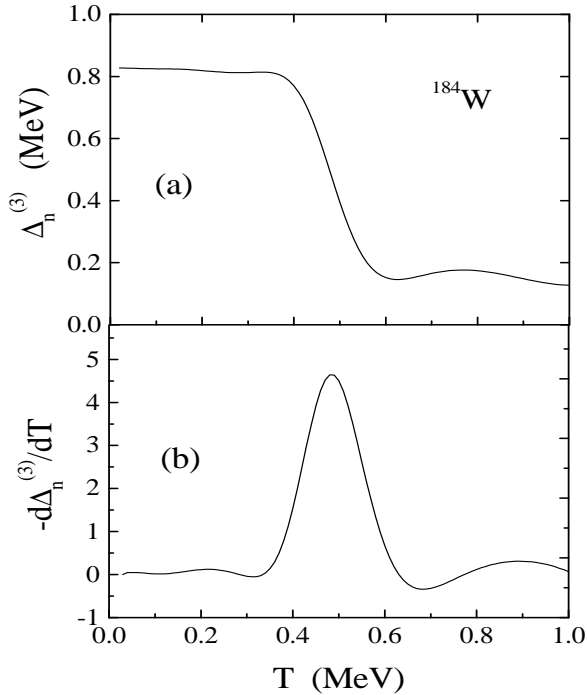}
  \caption{Thermal pairing gap and variation. (a)The thermal pairing gap (solid line) 
  extracted from the thermal odd-even mass 
  difference Eq. (\ref{eq:5}) as a function of temperature. 
  (b)The variation of the thermal odd-even mass difference defined by Eq. (\ref{eq:5}).}
  \label{fig3}
\end{figure}
\begin{figure}[b]
\includegraphics[width=8cm,height=10cm]{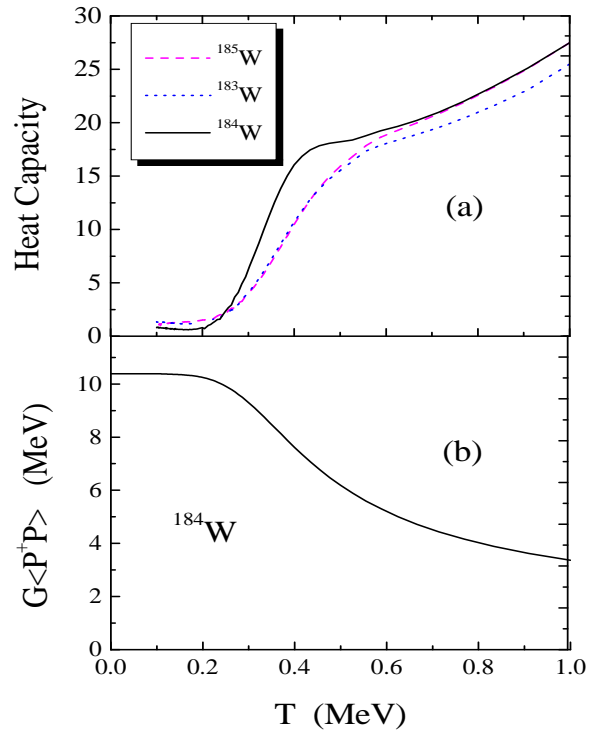}
  \caption{Calculated heat capacities and pairing gap as a function of temperature $T$ 
  in the SPA+RPA for the monopole pairing model. The upper graph (a) shows the heat 
  capacities where the broken, dotted, and 
  solid lines denote, respectively, those of $^{183}$W, $^{184}$W, and $^{185}$W. 
  The lower graph (b) shows the pairing energy. }
  \label{fig4}
\end{figure}

The heat capacities of $^{183}$W, $^{184}$W, and $^{185}$W are shown in Fig. \ref{fig2}. 
All the heat capacities exhibit the S shape with peaks around $T=$0.5 MeV. 
These heat capacities display characteristic behavior similar to those of $^{161,162}$Dy 
and $^{171,172}$Yb \cite{Schiller}. Moreover, we notice that the heat capacity of 
$^{184}$W around $T=$0.5 MeV is larger than those of $^{183}$W and $^{185}$W. 
The SPA + RPA \cite{Rossignoli} and SMMC \cite{Liu} calculations exhibited the S shape of the 
heat capacity and this odd-even effect where the heat capacity of an odd-mass nucleus is 
smaller than that of the adjacent even-even nuclei. 
We can see also the deviations from the heat capacity of the back-shifted Fermi gas model, 
which is approximated by the Bethe formula $C=2aT$. 
In the SMMC calculation \cite{Liu}, Liu and Alhassid identified a signature of the pairing 
transition in the heat capacity that is correlated with the reduction of 
the number of neutron pairs as the temperature 
increases. In their calculation, the pairing correlations are suppressed 
for even-even nuclei, but not for adjacent odd-mass nuclei. 

Figure \ref{fig3} (a) shows the thermal pairing gap extracted from the thermal odd-even mass 
difference defined by Eq. (\ref{eq:5}) as a function of temperature. 
We find a sudden decrease of the thermal odd-even mass difference curve around $T$= 0.5 MeV, 
which is interpreted as a rapid breaking 
of nucleon Cooper pairs and the suppression of pairing correlations. 
We can now regard an inflection point of the curve of $\Delta_{n}^{(3)}$ 
in Fig. \ref{fig3} (a) as a signature of pairing transition, and called it 
``transition temperature" in our previous paper \cite{Kaneko}. 
To see more precise position of the inflection point, we differentiate 
$\Delta_{n}^{(3)}(^{184}{\rm W})$ with respect to temperature $T$. 
It is very important to note that the thermal odd-even mass difference 
has the following identity: 
\begin{eqnarray}
-\frac{\partial\Delta_{n}^{(3)}(Z,N,T)}{\partial T}  =  (-1)^{N}\{ C(Z,N,T) \hspace{2cm} \nonumber \\
 \hspace{0cm} - \frac{1}{2}[C(Z,N+1,T) + C(Z,N-1,T)] \}. \nonumber \\
\label{eq:7}
\end{eqnarray}
This identity means that the odd-even difference in the heat capacities represents a variation 
of the pairing correlations depending on temperature. In Fig. \ref{fig3} (b), we can see that 
the peak of the odd-even difference in the heat capacities is a signature of the pairing 
transition, and the thermal odd-even mass difference is a good indicator for the pairing 
correlations. 
The extrapolation of the level dencity to $\sim 40$ MeV 
affects the curve of $\Delta_n^{(3)}$ at high temperature 
in Fig. 3. However, the effects do not change 
the sudden decrease of the pairing gap $\Delta_n^{(3)}$ 
around $T=0.5$ MeV. 

To describe the above characteristic behavior of the heat capacity and the pairing gap 
extracted from the measured level densities of $^{183}$W, $^{184}$W, and $^{185}$W, 
we consider a monopole pairing Hamiltonian 
\begin{eqnarray}
H & = & \sum_{k}\varepsilon_{k}(c_{k}^{\dag}c_{k}+c_{\bar{k}}^{\dag}c_{\bar{k}})-GP^{\dag}P, 
\label{eq:8}
\end{eqnarray}
where $\varepsilon_{k}$ are the single-particle energies and $P$ is 
the pairing operator $P=\sum_{k}c_{\bar{k}}c_{k}$. 
By means of the SPA+RPA \cite{Rossignoli,Attias} based on the Hubbard-Stratonovich 
transformation \cite{Hubbard}, the canonical partition function 
with the number parity projection $P_{\sigma}=(1+\sigma {\rm e}^{i\pi N})/2$ 
\cite{Tuominen,Rossignoli} is given by
\begin{eqnarray}
Z_{\rm c}^{\sigma} & = & {\rm Tr}[P_{\sigma}{\rm e}^{-H/T}]_{\rm SPA+RPA} \nonumber \\
& = & \frac{2}{GT}\int_{0}^{\infty}\Delta d\Delta 
{\rm e}^{-\Delta^{2}/GT}Z_{\sigma}C_{\rm RPA}. 
\label{eq:9}
\end{eqnarray}
where $\sigma$ means the even or odd number parity. 
Here 
\begin{eqnarray}
Z_{\sigma} = \frac{1}{2}\prod_{k}{\rm e}^{-\gamma_{k}/T}
(1+{\rm e}^{-\lambda_{k}/T})^{2} \hspace{0.5cm}\nonumber \\ 
 \cdot [1+\sigma \prod_{k'}{\rm tanh}^{2}(\lambda_{k'}/T)], \nonumber \\
C_{\rm RPA} = \prod_{k}\frac{\omega_{k}{\rm sinh}[\lambda_{k}/T]}
{2\lambda_{k}{\rm sinh}[\omega_{k}/2T]},
\label{eq:10}
\end{eqnarray}
where $\lambda_{k}=\sqrt{\varepsilon'^{2}_{k}+\Delta^{2}}$, 
$\varepsilon'_{k}=\varepsilon_{k} - \mu - G/2$, and 
$\gamma_{k}=\varepsilon_{k} - \mu - \lambda_{k}$. 
The $\omega_{k}$ are the conventional thermal RPA energies. If the factor $C_{RPA}$ is 
neglected, the SPA partition function is obtained. 
Then the thermal energy can be calculated from $E=-\partial {\rm ln}Z_{\rm c}^{\sigma}/\partial (1/T)$. 
In this calculation, we use the single-particle energies $\varepsilon_{k}$ given by 
an axially deformed Woods-Saxon potential with spin-orbit interaction \cite{Cwoik}. 
The Woods-Saxon parameters are chosen so as to fit the experimental single-particle 
enegies extracted from the energy levels of odd nucleus $^{133}$Sn 
($^{132}$Sn core plus one neutron). 
The deformation takes into account effects of a quadrupole-quadrupole 
interaction in the mean-field approximation. 
The deformation parameter $\beta=0.23$ can be estimated from the experimental 
value $B(E2)=119.3$ W.u. in the even-even nucleus $^{184}$W. 
The 50 doubly degenerate single-particle energies are taken by assuming 
$^{132}$Sn core, and we fix the pairing force strength at $G=20/A$ so that 
the BCS pairing gap $\Delta_{BCS}$ reproduce the experimental odd-even mass difference 
$\Delta$ = 0.83 MeV for $^{184}$W at zero temperature. 
Figure \ref{fig4} (a) shows the calculated heat capacities for $^{183}$W, $^{184}$W, 
and $^{185}$W. We can see that the characteristic behavior of the extracted heat 
capacities in Fig. \ref{fig2} are well described. Then the neutron pairing energy 
calculated from 
$G\langle P^{\dag}P \rangle = GT\partial {\rm ln}Z_{c}^{\sigma}/\partial G$
is also shown in Fig. \ref{fig4} (b). 
We notice that the calculated heat capacity of $^{183}$W deviates from 
that of $^{184}$W, in contrast with that of $^{185}$W, which is different 
from the result of Fig. 2. However, for high temperature this deviation would approach 
to zero according to Eq. (7) because the derivative of pairing gap 
$-\frac{\partial\Delta_{n}^{(3)}}{\partial T}$ converges to zero when $T \rightarrow \infty$. 
The monopole pairing model qualitatively explains 
the reduction of the thermal odd-even mass difference in Fig. \ref{fig3} (a). 
As pointed out in our previous paper \cite{Kaneko}, the peaks of the S-shaped heat capacities 
is quite close to the critical temperature $T_{c}= 0.57\Delta$ in the BCS theory. 

In conclusion, we have extracted the pairing gap of $^{184}$W at finite temperature 
from the experimental level densities of $^{183}$W, $^{184}$W, and $^{185}$W using 
thermal odd-even mass difference. The extracted pairing gap exhibits a similar behavior to 
that of in previous theoretical model predictions, that is, the pairing gap decreases 
from the value at zero temperature with increasing temperature. 
The calculations show that while there is no sharp phase transition, 
the pairing gap decreases with increasing temperature. 
In particluar, it decreases rapidly around the BCS critical temperature. Thus, we can 
demonstrate that the thermal odd-even mass difference is a good 
indicator for the pairing transition at finite temperature as well as usual one at zero 
temperature. 
In this paper, the monopole pairing model was used with the SPA+RPA to describe 
the heat capacity and the pairing gap, where the deformation effect was taken into 
account. For these quantities, 
however, the fluctuations \cite{Rossignoli} to the contributions 
of the quadrupole-quadrupole interaction should be taken into account in 
more realistic calculations. Further investigations are in progress. 
We suggest that the pairing correlations can be estimated from the measured 
level densities of the triplet nuclei with neutron number $N+1, N,$ and $N-1$. 
We hope for further experiments to extract the pairing gap at finite temperature. 

One of the authors (K.K.) thanks Dr. A. M. Sukhovoj for information of the experimental 
data and inspiring discussion. 




\begin{thebibliography} {99}

\bibitem{BCS} J. Bardeen, L. N. Cooper, and J. R. Schrieffer, Phys. Rev. {\bf 108}, 
              1175(1957).
\bibitem{Bohr} A. Bohr, B. R. Mottelson, and D. Pines, Phys. Rev. {\bf 110}, 936(1958) 
\bibitem{Belyaev} S. T. Belyaev, Mat. Fys. Medd. Dan. Vid. Selsk. {\bf 31}, No. 11(1959).
\bibitem{Sano} M. Sano and S. Yamasaki, Prog. Theor. Phys. {\bf 29}, 397(1963).
\bibitem{Goodman} A.L. Goodman, Nucl. Phys. {\bf A352}, 45(1981).
\bibitem{Schiller} A. Schiller, A. Bjerve, M. Guttormsen, M. Hjorth-Jensen, F. Ingebretsen, 
                   E. Melby, S. Messelt, J. Rekstad, S. Siem, and S. W. $\O$degard, Phys. 
                   Rev. C {\bf 63}, 021306(R)(2001).
\bibitem{Melby} E. Melby, L. Bergholt, M. Guttormsen, M. Hjorth-Jensen, F. Ingebretsen, 
                S. Messelt, J. Rekstad, A. Schiller, S. Siem, and S. W. $\O$degard, 
                Phys. Rev. Lett. {\bf 83}, 3150(1999). 
\bibitem{Rossignoli} R. Rossignoli, N. Canosa, and P. Ring, Phys. Rev. Lett. {\bf 80}, 
                     1853(1998).
\bibitem{Rombouts} S. Rombouts, K. Heyde, and N. Jachowicz, Phys. Rev. C {\bf 58}, 3295(1998).
\bibitem{Liu} S. Liu and Y. Alhassid, Phys. Rev. Lett. {\bf 87}, 022501(2001); 
             Y. Alhassid, G. F. Bertsch, and L. Fang, Phys. Rev. C {\bf 68}, 044322(2003). 
\bibitem{Egido} J. L. Egido, L. M. Robledo, and V. Martin, Phys. Rev. Lett. {\bf 85}, 26(2000). 
\bibitem{Agrawal} B. K. Agrawal, Taps Sil, S. K. Samaddar, and J. N. De, Phys. Rev. C 
                  {\bf 63}, 024002(2001). 
\bibitem{Tuominen} M. T. Tuominen, J. M. Hergenrother, T. S. Tighe, and M. Tinkham, Phys. Rev. 
                 Lett. {\bf 69},1997,(1992); 
                 P. Lafarge, P. Joyez, D. Esteve, C. Urbina, and M. H. Devoret, Phys. Rev. 
                  Lett. {\bf 70}, 994(1993). 
\bibitem{Kaneko} K. Kaneko and M. Hasegawa, Nucl. Phys. A {\bf 740}, 95(2004). 
\bibitem{Bondarenko} V. A. Bondarenko, J. Hanzatko, V. A. Khitrov, Li Chol, Yu, E. Loginov, 
                     S. Eh. Malyutenkova, A. M. Sukhovoj, and I. Tomandl, 
                     Report at XII International Seminar on Interaction of Neutrons 
                     with Nuclei, May 26-29 2004, Dubna, Russia.; nucl-ex/040630. 
\bibitem{Sukhovoj} A. M. Sukhovoj, private communication. 
\bibitem{Gilbert} A. Gilbert and A. G. W. Vameron, Can. J. Phys. {\bf 43},1446(1965); 
          T. von Egidy, H. H. Schmidt, and A. N. Behkami, Nucl. Phys. {\bf A481}, 189(1988). 
\bibitem{Hubbard} J. Hubbard, Phys. Rev. Lett. {\bf 3}, 77(1959); 
                  R. L. Stratonovich, Dokl. Akad. Nauk S.S.S.R. {\bf 115}, 1097(1957). 
\bibitem{Attias} H. Attias and Y. Alhassid, Nucl. Phys. {\bf A625}, 565(1997). 
\bibitem{Cwoik} S. Cwoik, J. Dudek, W. Nazarewicz, J. Skalski, and T. Werner, Comput. Phys. 
                Commun. {\bf 46}, 379(1987). 
\end{thebibliography}
\end{document}